\documentclass[pra,twocolumn,amssymb,showpacs]{revtex4}

\usepackage{graphicx}
\usepackage{appendix}

\begin{document}

\title{Critical dynamics of decoherence}
\author{Bogdan Damski, H. T. Quan, and Wojciech H. Zurek}
\affiliation{Theoretical Division, MS B213, Los Alamos National Laboratory, Los Alamos,
NM, 87545, USA}
\begin{abstract}
 We study
decoherence induced by a dynamic environment undergoing a quantum phase transition.
Environment's susceptibility  to perturbations
-- and, consequently, efficiency of decoherence -- is 
amplified near a critical point. 
Over and above this near-critical susceptibility increase, we show that decoherence is 
dramatically enhanced by  non-equilibrium critical dynamics of the environment.
We derive a simple expression relating decoherence to the universal 
critical exponents exhibiting deep connections with the 
theory of topological defect creation 
in non-equilibrium phase transitions.
\end{abstract}
\maketitle

\section{Introduction}

Decoherence of a qubit coupled to an environment is studied  with the
 central spin models.  
In addition to being a workhorse 
for understanding of quantum to classical transition \cite{1decoh,2decoh,Zurek81}, 
central spin models 
describe loss of qubit's coherence in 
NV centers in diamond \cite{HansonScience},  quantum dots in semiconductors \cite{CentralSpin}, 
NMR experiments \cite{JingfuPRL,FercookNMR}, etc. 
The environment, 
usually regarded as a destroyer of quantum coherence and entanglement, has been  recently
proposed to enhance sensitivity of a  magnetometer based on the central spin system 
\cite{Goldstein2010}. Therefore, understanding how the environment influences 
qubit's state is central to quantum computing and metrology.

So far, studies of decoherence focused mainly on systems prepared in Schr\"odinger cat-like 
superpositions instantaneously coupled to an 
otherwise static environment described by time-independent Hamiltonian.  
We consider instead a spin-1/2 system
 coupled to a driven    environment. The environment 
undergoes  a quantum phase transition (QPT) as it is driven  
through the quantum critical point  by the change of the external field
(e.g., a magnetic field in the Ising chain depicted in  Fig. \ref{fig1}).
Decoherence occurs because the environment ``monitors'' the system 
and ``finds out'' its pointer states (i.e., effectively classical states 
preserved in spite of  decoherence \cite{1decoh,2decoh,Zurek81}). This means that each pointer 
state remains untouched, but makes a distinct imprint on the quantum state 
of the environment. 
How close to orthogonal are these record states determines how well the 
environment  knows about the system. 
Their overlap is the decoherence factor $d(t)$ multiplying 
off-diagonal terms of the reduced density matrix of the qubit.

Effectiveness of decoherence depends on how easy it is 
to perturb -- ``write on'' --  
the environment: 
Its ``stiffness'' is determined by its Hamiltonian as well as by its state. 
For the Ising environment, Hamiltonian-induced 
stiffness is due to either spin-spin interactions or 
the external field (Fig. \ref{fig1}). When the two balance, the environment 
is at the critical point, where it is most susceptible to perturbations. 
Decoherence is expected to be strongest there. This was already seen 
in {\it static} critical environments \cite{quan06,paz}.
In our system  the environment is {\it not} static: It is driven across the phase 
transition, so its instantaneous state -- and thus decoherence --  is 
determined by both its Hamiltonian and its history. We show that the latter
profoundly affects qubit's decoherence rate.

The aim of this paper is to understand how a driven critical environment 
affects decoherence of a  qubit. 
For clarity of our discussion, which can be extended to more complicated systems, 
we consider an exactly solvable paradigmatic 
model of a QPT: the Ising chain coupled to a central spin-1/2
(Fig. \ref{fig1}). 
We show analytically that  decoherence is enhanced by the 
nonequilibrium transition and exhibit interesting dynamics characterized 
by the transition rate, coupling strengths and ground state fidelity. 
We  also show that 
the time-dependent decoherence factor encodes 
both  
universal properties of the system associated with quantum criticality
(critical exponents) 
and rich system-dependent (non-universal) features.  
Our work combines theory of decoherence with rapidly growing fields of 
dynamics of QPTs 
\cite{BD2005,dorner,polkovnikov,Jacek2005,DziarmagaReview,1quench_dyn_exp}
and quantum information \cite{Zanardi2006,GuReview,osterloh,MMR2010}. 
In the latter context, our study can be regarded as a dynamic generalization of
the so-called fidelity approach to QPTs \cite{Zanardi2006,GuReview}. We would also like to 
mention that in the
context of dynamics of QPTs decoherence has been studied in
\cite{Ralf_decoh,Santoro_decoh,Marek_decoh}.

\section{Basics of quench dynamics} 
To study non-equilibrium environment -- qubit evolution we  characterize dynamics 
of an environment driven across a critical point. 
This is done using  the quantum version  \cite{dorner,polkovnikov,BD2005}
of the Kibble-Zurek (KZ) theory \cite{1kzm,Znature}. 
The environment is quenched by decreasing the
external field 
\begin{equation}
g(t)=g_c-t/\tau_{Q},
\label{g_tt}
\end{equation}
where $g_c$ stands for strength of the external field at the critical point, 
the quench time $\tau_{Q}$ is inversely proportional to the quench  
speed, and time $t$ goes from $-\infty$ to $+\infty$.
A QPT is characterized by 
vanishing excitation gap $\Delta_0|g-g_c|^{z\nu}$ and divergent correlation 
length $\xi_0/|g-g_c|^\nu$, where $z$ and $\nu$ are the critical exponents, 
while $\Delta_0$ and $\xi_0$ are constants \cite{sachdev}. There are two relevant
time scales during a quench: the system reaction time given by 
$\hbar/\Delta_0|g-g_c|^{z\nu}$ and the time scale of change of the Hamiltonian 
given by 
$(g(t)-g_c)/\frac{d}{dt}(g(t)-g_c)$.
Away from the critical point  the reaction time is short compared to quench time scale 
and so  evolution is adiabatic. Near the critical point  opposite happens and 
 evolution is  approximately diabatic. 
The border between the two regimes lies at a distance $\hat g$ from 
the critical point. Equating the two timescales, we get  
\begin{equation}
\hat g\sim\tau_Q^{-1/(1+z\nu)}.
\label{gihat}
\end{equation}
KZ theory also predicts that there is a non-equilibrium length scale $\hat\xi$
imprinted
onto the post-transition state of the system \cite{Znature}. 
It is given by the coherence length at the border between 
adiabatic and impulse regimes where the state of the many-body system freezes out:
\begin{equation}
\hat\xi = \xi(\hat g)\sim\tau_Q^{\nu/(1+z\nu)}.
\label{hat_xi}
\end{equation}
This scale yields e.g. a typical distance between topological defects
created during a symmetry-breaking QPT. It also suggests that 
non-equilibrium excitations resulting from the quench 
should encode a characteristic momentum scale
\begin{equation}
\hat k\sim \hat\xi^{-1}\sim\tau_Q^{-\nu/(1+z\nu)}.
\label{khat}
\end{equation}

\section{The Model}
\label{Sec_Model}
\begin{figure}
\includegraphics[width=\columnwidth, clip=true]{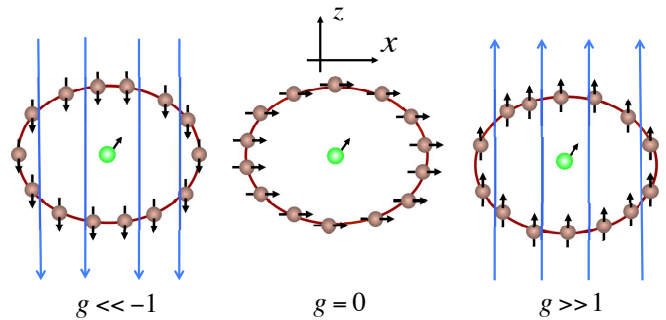}
\caption{(color online).
The  central spin-1/2 (qubit)  
interacts with the Ising spin environment 
whose quantum phase transition is driven by an external magnetic field.
For large initial and final fields -- 
right and left  panel,  respectively -- the environment  is in the paramagnetic phase: 
its spins align with the field. 
For small fields ($|g(t)|<1$, center panel) Ising chain enters a ferromagnetic phase, 
where its spins try to align with $x$ or $-x$ breaking the symmetry of  
the Ising Hamiltonian. 
}
\label{fig1}
\end{figure}

We consider decoherence 
of the qubit coupled to a driven 
Ising chain undergoing a QPT (Fig \ref{fig1}).
The Ising  Hamiltonian is 
$$
\hat H_{\cal E}= -\sum_{j=1}^N \left(\sigma _{j}^{x}\sigma_{j+1}^{x}+g(t)\sigma _{j}^{z}\right),
$$
while its coupling to the qubit reads
$$
\hat H_{\cal SE}=-\delta   \sum_{j=1}^N\sigma _{j}^{z}\sigma^{z}_{\cal S}.
$$
Above the environment size $N\gg1$, the system-environment coupling $\delta\ll1$,
and $\sigma^{z}_{\cal S}$ is the Pauli matrix of the qubit. 
Ising chain in a transverse field 
has been  experimentally studied in \cite{ising_exp,JingfuPRA2009}. 
The coupling $\hat H_{\cal SE}$ has been implemented in
\cite{JingfuPRL}. The total qubit-environment Hamiltonian 
is given by 
$$
\hat H = \hat H_{\cal E} + \hat H_{\cal SE}.
$$
The driving is a result of a slow -- $\tau_Q\gg1$ -- ramp down of a magnetic field 
\begin{equation}
g(t)=1-t/\tau_Q,
\label{g_t}
\end{equation}
which is equivalent to (\ref{g_tt}) taken with $g_c=1$.

\noindent Evolution starts at $t=-\infty$ with the qubit  in a pure state  
superposition of $c_{+} \left |\uparrow \right \rangle +c_{-}  
\left |\downarrow \right\rangle $ and the environment  
in its instantaneous ground state $|GS(t=-\infty)\rangle$.
The composite  wave function $|\psi\rangle$ of the system is given initially
by 
$$
|\psi(t=-\infty)\rangle = (c_{+} \left |\uparrow \right \rangle +c_{-}  \left |\downarrow \right\rangle)
\otimes|GS(t=-\infty)\rangle.
$$
A straightforward calculation then shows that 
\begin{eqnarray*}
|\psi(t)\rangle &=& \hat Te^{-i\int_{-\infty}^t dt\,\hat H(g(t))}
|\psi(t=-\infty)\rangle\nonumber\\ &=&
c_+|\uparrow\rangle\otimes
\hat Te^{-i\int_{-\infty}^t dt\hat H_{\cal E}(g(t)+\delta)}
|GS(t=-\infty)\rangle \\
&+& 
c_-|\downarrow\rangle\otimes
\hat Te^{-i\int_{-\infty}^t dt\hat H_{\cal E}(g(t)-\delta)}
|GS(t=-\infty)\rangle \\
&=&  c_{+} \left| \uparrow \right\rangle \otimes | 
\varphi _{+}(t)\rangle+c_-\left| \downarrow \right\rangle \otimes |  
\varphi _{-}(t)\rangle,
\end{eqnarray*}
where $\hat T$ is the time-ordering operator, and evolution of the
environmental states coupled to up-down qubit states is given by 
\begin{equation}
i\frac{\partial} 
{\partial t} \left\vert
\varphi_\pm(t)
\right\rangle = \hat H_{\cal E}(g(t)\pm\delta)\left\vert
\varphi_\pm(t)\right\rangle.
\label{env_dt}
\end{equation}
Therefore, evolution of the system 
depends  on  the dynamics of two Ising branches 
evolving  in an effective magnetic field given by 
$g(t)\pm\delta$.

The reduced density matrix of the qubit in the
$\protect{\{|\uparrow\rangle,|\downarrow\rangle\}}$ basis reads
$$\rho_{\cal S}(t)= {\rm Tr}_{\cal E}|\psi(t)\rangle\langle\psi(t)|=
\left(
\begin{array}{cc}
|c_+|^{2} &   c_+ c_-^* d^*(t) \\
c_+^* c_- d(t) & |c_-|^{2}  
\end{array}
\right),
$$  
where $d(t)=\langle\varphi_+(t)|\varphi_-(t)\rangle$ is the decoherence factor.
We study its squared modulus, 
$$
D= |d(t)|^2 = |\langle\varphi_+(t)|\varphi_-(t)\rangle|^2,
$$
also known as the decoherence factor; we follow this nomenclature below.
When $D=1$, the qubit  is in a pure state, but when $D=0$ it
is completely decohered.

\section{Exact solution}
Dynamics of the decoherence factor can be obtained analytically 
by solving  (\ref{env_dt}). This can be done  after mapping the spins onto 
non-interacting fermions via the Jordan-Wigner transformation \cite{Jacek2005}.
We introduce after Dziarmaga  the Bogolubov
modes $u^\pm_k$ and $v^\pm_k$  through 
\begin{equation}
\left\vert \varphi _\pm(t) \right\rangle = 
\prod_{k>0} \left [ 
u^\pm_{k}(t) \left\vert 0_{k}, 0_{-k}\right\rangle
-v^\pm_{k}(t) \left\vert 1_{k}, 1_{-k}\right\rangle\right],
\label{waveK}
\end{equation}
where 
$\left\vert m_{k}, m_{-k}\right\rangle$ describes the state with $m=0,1$ pairs 
of quasiparticles with momentum $k=(2s+1)\pi/N$,  
$s=0,1,...,N/2-1$. We consider even $N$. 
In this formalism one easily finds that 
\begin{equation}
D(t)= \prod_{k>0} F_k(t), \ F_k(t) = \left|u_{k}^{+*}(t)u _{k}^-(t) + v_{k}^{+*}(t) v_{k}^-(t)\right|^2.
\label{Dprod}
\end{equation}
We will refer to $F_k$ to describe dynamics of decoherence from the
momentum space angle.
We call it  a decoherence factor in momentum space.
Using (\ref{Dprod}) we find that for thermodynamically large environments
\begin{equation}
D(t) \approx \exp\left(-\frac{N}{2\pi}\int_0^\pi dk \ln
F^{-1}_k(t)\right).
\label{LB}
\end{equation}
The above integral representation shall be accurate for time instants $t$  for which 
$\ln F_k(t)$ is non-singular $\forall k$ \cite{Disc_Inte}.
All we need now are the Bogolubov modes at various stages of 
evolution. The modes evolve according to  \cite{Jacek2005}
\begin{eqnarray*}
i\frac{d}{dt}v^{\pm}_{k}&=& -2(g(t)\pm\delta-\cos k)v^{\pm}_{k} + 2\sin ku^{\pm}_k,\\
i\frac{d}{dt} u^{\pm}_{k}&=& 2 (g(t)\pm\delta-\cos k) u^\pm_k + 2\sin k v^{\pm}_{k}.
\end{eqnarray*}
We apply the transformation 
$t'_\pm=4\tau_Q\sin k\left(-g(t)\mp\delta+\cos k\right)$.
Defining $\tau_Q'=4\tau_Q\sin^2k$ we end up with the Landau-Zener
system, 
\begin{equation}
i\frac{d}{dt'_\pm}
\left(
\begin{array}{c}
v^{\pm}_{k} \\
u^{\pm}_{k}
\end{array}
\right)
= 
\frac{1}{2}\left(
\begin{array}{lr}
\frac{t'_\pm}{\tau_Q'} & 1 \\
1  & -\frac{t'_\pm}{\tau_Q'} 
\end{array}
\right)
\left(
\begin{array}{c}
v^{\pm}_{k} \\
u^{\pm}_{k}
\end{array}
\right),
\label{grypa}
\end{equation}
that can be solved with the Weber functions (see e.g. \cite{BDZ2006}). The initial conditions 
are $v_k^\pm(t'_\pm=-\infty)=0$ and $u_k^\pm(t'_\pm=-\infty)=1$.
To express the solution in a compact form we
introduce:
$z_\pm = t'_\pm \exp(i\pi/4)/\sqrt{\tau_Q'}$ and $n = -i\tau_Q'/4$.
The exact solution reads
\begin{eqnarray}
\label{exact}
&v_k^\pm(t)& = \frac{\sqrt{\tau_Q'}}{2}
\exp\left(-\pi\tau_Q'/16\right){\cal D}_{-n-1}(iz_\pm)\\
&u_k^\pm(t)& = \exp\left(-\pi\tau_Q'/16\right)\exp(i\pi/4)\times \nonumber\\ &&
[(1+n)
{\cal D}_{-n-2}(iz_\pm) +iz_\pm {\cal D}_{-n-1}(iz_\pm)]
\label{exact1}
\end{eqnarray}
where ${\cal D}_m(iz)$ is the Weber function \cite{Whittaker}.
It is simplified in the Appendix.

\section{Dynamics of decoherence}
\label{fizyka}
\begin{figure}
\includegraphics[width=\columnwidth, clip]{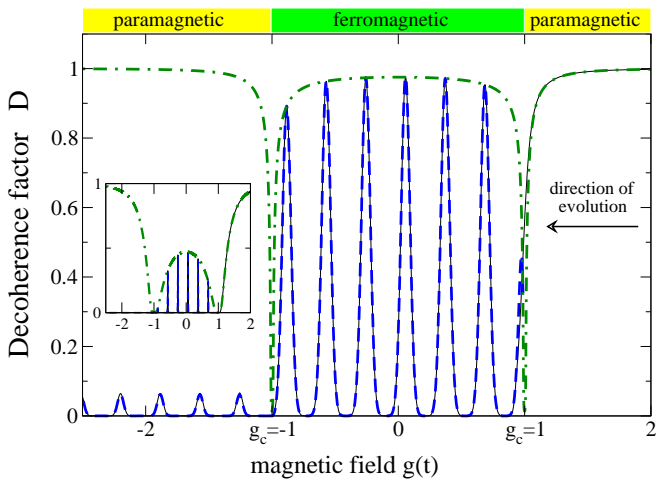}
\caption{(color online) Decoherence  during a quench: a quasi-periodic regime between 
the critical points.  Black solid line shows the result obtained from  
numerical integration of (\ref{grypa}).
Our analytical approximations are superimposed on it as a blue  dashed
line: (\ref{L_approx}) for $g(t)\in(-1,1)$ and (\ref{L_approx1}) for $g(t)<-1$. The 
green dashed-dotted  line is the adiabatic result equivalent to 
ground state fidelity ($\tau_Q\to\infty$ limit): (\ref{swinka}) combined with 
(\ref{Dprod}). 
Numerical solution 
almost perfectly overlaps with analytical results away from the critical
points: $|g(t)-g_c|>\hat g,\delta$. 
The main plot shows that for moderate environment sizes -- $N=1000$ here -- 
almost perfect revivals of coherence take place 
between the critical points, but  coherence is virtually lost 
after driving the environment through the second critical point.
The inset shows the same, but for $N=30,000$. It illustrates that for very 
large environments there
is little coherence left after passing the first critical point
and the qubit is completely decohered after crossing  
the second critical point.
Note that ground state fidelity provides the envelope for the 
revivals of coherence between the critical points.
We used here  $\delta=10^{-2}$ and $\tau_Q=250$.
}
\label{fig2}
\end{figure}

\begin{figure}
\includegraphics[width=\columnwidth, clip]{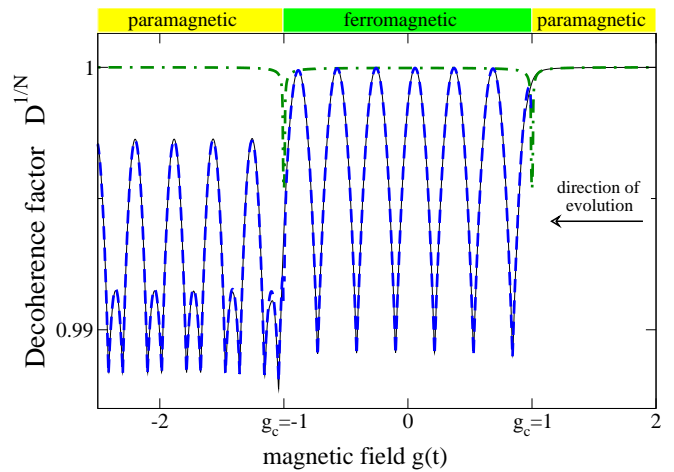}
\caption{(color online) Same as in Fig. \ref{fig2}, except we consider here
N-th root of  the decoherence factor $D$ 
illustrating  details of dynamics of decoherence  in
the large $N$ limit. Numerical result (black line) is obtained by calculating $D^{1/N}$ for $N=2000$ environmental spins.
$N$-th root of ground state fidelity (green dashed-dotted line) is calculated for the same $N$.
}
\label{fig3}
\end{figure}

\begin{figure}
\includegraphics[width=\columnwidth, clip]{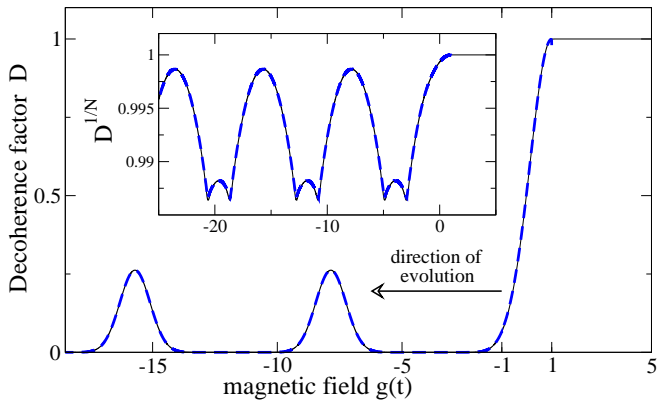}
\caption{(color online) Decoherence  during a quench: a gaussian regime
between the critical points.
Solid black line shows the  result coming from 
numerical integration of (\ref{grypa}).
The dashed blue line shows our analytical approximations: 
the gaussian result (\ref{gauss}) between the critical points
[$g(t)\in(-1,1)$] and  (\ref{L_approx1}) past the second critical 
point [$g(t)<-1$]. 
The inset shows $N$-th root of the data from the main plot: it illustrates
details of the revivals of coherence past the second critical point.
The parameters for this plot are: $N=1000$, $\delta=4\times10^{-4}$, 
and $\tau_Q=250$. Except for $\delta$, they are the same as in 
Fig. \ref{fig2}. The switch from coherence revivals (Fig. \ref{fig2}) to monotonic 
decay of coherence  between the critical points is a result of lowering the 
qubit -- environment coupling $\delta$ below the $\pi/16\tau_Q$
threshold. 
}
\label{fig4}
\end{figure}

The quench starts with the environment in the 
paramagnetic phase ($g(t)>1$), moves the environment across the ferromagnetic
phase ($-1<g(t)<1$), and finally brings it to the ``other'' paramagnetic phase ($g(t)<-1$).
The critical points are located at  $g_c=\pm1$. 
Substituting  $z,\nu=1$ into (\ref{gihat}) and (\ref{hat_xi}) results
in $\hat g\sim 1/\sqrt{\tau_Q}$ and $\hat \xi\sim\sqrt{\tau_Q}$ \cite{dorner}. 
The non-equilibrium length scale $\hat\xi$ 
provides the typical spacing of kinks (spin flips) excited during the transition \cite{dorner,Jacek2005}. 
Dynamics of the decoherence factor is depicted in Figs. \ref{fig2}-\ref{fig4}.

\begin{figure}
\includegraphics[width=\columnwidth, clip]{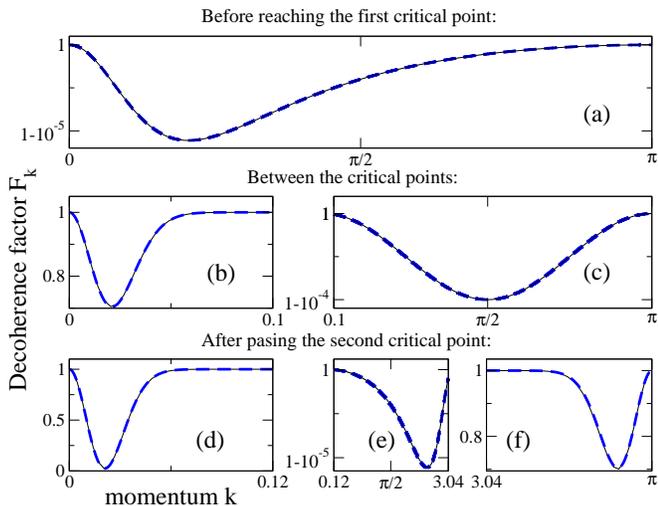}
\caption{(color online) Decoherence factor $F_k$  during the quench.
Solid black lines show numerics based on (\ref{grypa}).
The blue dashed lines  illustrate our analytical approximations.
Before reaching the critical point at $g_c=1$ modes evolve adiabatically:
panel (a) made for $g(t)=2$. The adiabatic result (\ref{ad_fk}) 
 fits numerics for all momenta. 
 Between the critical points -- panels (b) and (c) made for $g(t)=0$ -- 
 low $k$ modes are excited due to driving across $g_c=1$, while the large $k$ modes
 evolve adiabatically. The former is illustrated in panel (b) where 
 (\ref{fk_approx}) fits numerics, while the latter is depicted in panel (c) 
 where (\ref{ad_fk}) matches numerics.
 Panels (d-f) prepared for $g(t)=-2$ show what happens  
 after passing the second critical point at $g_c=-1$:
  low $k$ modes are still
 excited due to driving across $g_c=1$, 
 intermediate modes evolve adiabatically, while large $k$ modes are 
 excited by crossing the $g_c=-1$ critical point. This is 
 theoretically described by (\ref{fk_approx}), (\ref{ad_fk}) and 
 (\ref{fk_approx1}) depicted in panels (d), (e) and (f), respectively.
Note that despite the tiny range of variation
of the adiabatically evolving  modes -- panels (a), (c) and (e) --
the striking envelope of the quasi-periodic revivals of 
coherence between the critical points comes from these modes.
Solid black lines (numerics) have been obtained for $N=2000$,
$\tau_Q=250$ and $\delta=10^{-2}$. The same $\tau_Q$ and $\delta$
have been used to get the dashed lines (analytical approximations). 
}
\label{fig5}
\end{figure}

As the field polarizing spins in the chain is ramped down (\ref{g_t}),
environment's sensitivity to external influence increases.
This results in enhanced decoherence visible as a gradual decrease of
the decoherence factor $D$ away from the critical point (Fig. \ref{fig2}). 
Initially evolution 
proceeds adiabatically and $D$
is identical to ground state fidelity of the environment \cite{Zanardi2006,GuReview,MMR2010}.
Namely, to the squared overlap between the ground states of $\hat H_{\cal E}(g\pm\delta)$:
$|\langle GS(g+\delta)|GS(g-\delta)\rangle|^2$. 
As is derived in the Appendix,  away from the critical points, $|g(t)-g_c|\gg\delta$, 
we have  for the $k$ modes evolving adiabatically 
\begin{equation}
F_k(t)=1-\frac{\delta^2\sin^2k}{(1-2g(t)\cos k+g(t)^2)^2}+{\cal
O}(\delta^4),
\label{ad_fk}
\end{equation}
which  compares well to numerics (Fig. \ref{fig5}a; see also Figs. \ref{fig5}c
and \ref{fig5}e).
Using (\ref{LB}) we obtain 
in the paramagnetic phase that 
\begin{equation}
D(t)\approx\exp\left(-\frac{N\delta^2}{4g(t)^2(g(t)^2-1)}\right).
\label{Dfidel}
\end{equation}

\noindent 
Just above the first critical point, $1<g< 1+\hat g$,  the adiabatic 
approximation breaks down and decoherence speeds up; decoherence
 factor decreases substantially.
This happens because (i) environment's sensitivity is enhanced by quantum criticality
amplifying the perturbation resulting from the coupling 
to the qubit system; (ii) the environment  is no longer in the ground state:
quench is taking it into a superposition of energy eigenstates.

As the environment is driven past the first critical point at $g_c=1$, 
there are either partial revivals
 (Figs. \ref{fig2} and \ref{fig3}) or a monotonic  decay
(Fig. \ref{fig4}) of coherence between the critical points. 
They  result from non-equilibrium 
dynamics around  the critical point  leading to excitation of small $k$ modes (the gap 
in the excitation spectrum closes for $k=0$ at $g_c=1$).
Expanding  (\ref{exact}) and (\ref{exact1}) -- see the Appendix -- we find 
a remarkably simple and
accurate solution for the problem.
For low $k$ modes, i.e., $k\sim \hat k={\cal O}(1)/\sqrt{\tau_Q}$ (\ref{khat}), 
we get
\begin{equation}
F_k(t)
=1-4\left(e^{-2\pi\tau_Qk^2}-e^{-4\pi\tau_Qk^2}\right)\sin^2\phi(t),
\label{fk_approx}
\end{equation}
where $\phi(t)\approx4(1-g(t))\tau_Q\delta=4t\delta$. As shown in Fig.
\ref{fig5}b -- see also Fig. \ref{fig5}d -- there is a very good agreement 
between (\ref{fk_approx}) and numerics.
The larger $k$ modes, those for which $k\gg\hat k$, evolve adiabatically through the 
critical point. Their  decoherence factor  $F_k$ reads (\ref{ad_fk}), which is 
illustrated in Fig. \ref{fig5}c, where again numerics matches well our  theory. 

\noindent One striking feature of (\ref{fk_approx}) is that there is a
characteristic momentum scale imprinted by the quench. It is   predicted
by the KZ theory (\ref{khat}). Indeed, (\ref{fk_approx}) depends on momentum through the combination of $k/\hat k$.
Moreover, the modes centered around 
$$
k_m = \sqrt{\frac{\ln 2}{2 \pi}} \frac{1}{\sqrt{\tau_{Q}}}
$$
contribute most to non-equilibrium decoherence resulting from crossing the 
first critical point: (\ref{fk_approx}) has a single minimum at $k_m$. Interestingly, the position of this 
minimum is fixed in the post-transition state, $dk_m/dt=0$ for $g(t)<1$. It   
means that the characteristic length scale 
imprinted by  the non-equilibrium critical dynamics remains ``frozen'' 
 after the critical point was passed. This is in perfect agreement with
the adiabatic-impulse picture of dynamics proposed by the quantum KZ theory 
\cite{dorner,BD2005}.

Combining (\ref{ad_fk}), (\ref{fk_approx}),  and (\ref{LB})  we get  
\begin{equation}
D(t)\approx \exp\left(-\frac{N\delta^2}{4(1-g(t)^2)}\right) \exp\left(-\frac{N}{2\pi} \frac{f[\phi(t)]}{\sqrt{\tau_Q}}\right),
\label{L_approx}
\end{equation}
where $f[\phi]=-\frac{1}{\sqrt{2\pi}}\int_0^\infty ds 
\ln[1-4(e^{-s^2}-e^{-2s^2})\sin^2\phi]$ and $\phi\approx4 t\delta$ 
(the $\tau_Q\to\infty$ limit was assumed to simplify the 
upper integration limit in $f[\phi]$) \cite{Disc_Inte}.
The first factor  above is again ground state fidelity, but 
this time in the ferromagnetic phase: notice the lack of $g^{-2}$ factor
as compared to the paramagnetic phase result  (\ref{Dfidel}).
The second
one comes from non-equilibrium environment excitation   
happening around the first critical point.
Indeed, since we already know that the quench enforces  $F_k=F(k/\hat k)$ for
low momentum modes, a simple change of the 
integration variable to $k/\hat k$ in (\ref{LB}) 
gives the $\tau_Q^{-\nu/(1+z\nu)}=\tau_Q^{-1/2}$ factor in (\ref{L_approx}); see also Sec. \ref{sec_general}. 
Thus, we show analytically that for the Ising chain  non-equilibrium 
contribution to the decoherence rate  
encodes {\it universal} information about 
the criticality of the environment via the $z$ and $\nu$ exponents. 
This information 
is recorded in the decohering state of the simplest quantum 
system, the spin-1/2 qubit, and  should be experimentally 
accessible.

The non-universal contribution to non-equilibrium decoherence, 
$f[\phi(t)]$ from (\ref{L_approx}), leads to  
rich dynamics of decoherence. We observe partial revivals of coherence 
between the critical points taking place when 
the environment-qubit coupling is strong enough: $\delta>\pi/16\tau_Q$.
They happen in the magnetic field ($g$) domain with the period of $\pi/4\tau_Q\delta$. 
The envelope of these revivals is given accurately by ground state 
fidelity: at the peaks of the decoherence factor non-adiabatic $F_k$'s (\ref{fk_approx}) 
equal unity and so only adiabatic modes  contribute to the decoherence factor 
there. As ground state fidelity  is smaller than unity, the revivals of coherence cannot be perfect (Fig. \ref{fig2}). 
The presence of the adiabatic
``fidelity'' envelope, however, may allow for {\it measurement} of ground state
fidelity at the peaks of the decoherence factor: an interesting
observation as we have here an out of equilibrium 
system and ground state fidelity is an equilibrium quantity.

In the opposite limit of weak coupling between 
the environment and the qubit, 
$\delta<\pi/16\tau_Q$, decoherence factor decreases monotonically 
till the second critical point at $g_c=-1$. In particular, when $\delta\ll\pi/16\tau_Q$ 
expansion of (\ref{L_approx}) gives  gaussian decay of $D$ (Fig. \ref{fig4}): 
\begin{equation}
D(t)\approx\exp\left(\frac{-N8\delta^2t^2(\sqrt{2}-1)}{\pi\sqrt{\tau_Q}}\right)\exp\left(\frac{-N\delta^2}{4(1-g(t)^2)}\right).
\label{gauss}
\end{equation}

Even richer dynamics of decoherence emerges
when the magnetic field reverses polarity and drives the system  
past the second critical point at $g_c=-1$ (Figs. \ref{fig2}-\ref{fig4}). 
Indeed, at the second critical point the modes 
$k\sim \pi-\hat k$ become excited because  the gap closes
there at  $k=\pi$. This means 
that the decoherence factor $F_k$  will be given by (\ref{fk_approx})
for $k\sim\hat k$, (\ref{ad_fk}) 
for $\hat  k  \lesssim k \lesssim \pi -\hat k$, and 
\begin{equation}
F_k(t) =1-4
\left(e^{-2\pi\tau_Q(k-\pi)^2}-e^{-4\pi\tau_Q(k-\pi)^2}\right)\sin^2\eta(t),
\label{fk_approx1}
\end{equation}
for $k\sim\pi-\hat k$, where $\eta(t)\approx4(g(t)+1)\tau_Q\delta$ (see the Appendix).
This is illustrated in Figs. \ref{fig5}d-f, where good agreement between
theory and numerics is confirmed. Similarily as above, the momentum
scale from the KZ theory leaves a long lasting imprint here: the modes most 
contributing to non-equilibrium decoherence due to crossing of the second critical
point are centered around $k=\pi-k_m$ for $g(t)<-1$: (\ref{fk_approx1}) has a
minimum at such $k$.

Marrying (\ref{ad_fk}), (\ref{fk_approx}) and (\ref{fk_approx1}) with (\ref{LB}) we obtain \cite{Disc_Inte}
\begin{eqnarray}
D(t)&\approx& \exp\left(-\frac{N\delta^2}{4g(t)^2(g(t)^2-1)}\right)\times
\nonumber \\
&&\exp\left(-\frac{N}{2\pi} \frac{f[\phi(t)] + f[\eta(t)]}{\sqrt{\tau_Q}}\right).
\label{L_approx1}
\end{eqnarray}
Comparing (\ref{L_approx1}) to (\ref{L_approx}) 
we see one additional
factor, $\exp(-Nf[\eta(t)]/2\pi\sqrt{\tau_Q})$, due to non-equilibrium transition
across the second critical point. We would like to stress again that 
dynamics across the two critical points happens ``independently'' because 
the gap of the environment 
closes at  $g_c=1$ ($g_c=-1$) for $k_0=0$ ($k_0=\pi$):
the first (second) transition excites small (large) momentum modes.
Excitations created around the two critical points
impose the same revival period for $D$, 
but their phases are shifted. 
It leads to destructive interference damping  the amplitude 
of the partial revivals  dramatically. 
We also note again that  (\ref{ad_fk}-\ref{L_approx1}) 
are derived away from the critical points, i.e., for 
$|g(t)-g_c|\gg\delta,\hat g$.

Finally, we compare the quench-amplified decoherence factor to the one resulting from 
the adiabatic quench ($\tau_Q=\infty$). We see in Figs. \ref{fig2} and \ref{fig3}
that decoherence can be 
much stronger in the driven environment: compare the green dashed-dotted line to the blue dashed line.
This is related to the excitations (kinks for the Ising chain \cite{dorner,Jacek2005}) 
appearing  in a non-adiabatic quench. 
Thus, susceptibility of many-body environments to perturbations reflects 
not just their instantaneous Hamiltonians 
or eigenstates, 
but also their  actual (non-equilibrium)
 state encoding the quench history. 

\section{General scaling result}
\label{sec_general}

A variety of systems have the decoherence factor $D$ given in the product form (\ref{Dprod}). 
This factorization  appears when a many-body 
environment is described by a product wave function 
in momentum space such as (\ref{waveK}). This is the case for  some
quantum magnets (XY, Ising, Kitaev, etc.) \cite{DziarmagaReview}, 
conventional superconductors  (either electronic \cite{Tinkham} or those emulated by
cold atoms \cite{Ketterle}), and liquid $^3$He \cite{He3}.
As for the Ising case, 
excitation of these environments during a dynamical phase transition 
shall imprint a non-equilibrium length scale $\hat\xi$ and 
momentum scale $\hat k$, (\ref{hat_xi}) and (\ref{khat}), respectively.
Note that we again 
assume that the external field   driving the QPT is quenched according to (\ref{g_tt}).
It is thus natural to assume that for those systems  as well overlap $F_k$ 
 between dynamically excited states of the environment 
depends on momentum through $(k-k_0)/\hat k$ combination:
$F_k=F((k-k_0)/\hat k)$, where $k_0$ is the momentum for 
which the gap closes. When that happens, exchanging the product (\ref{Dprod}) into an integral 
similarly as in (\ref{LB}), one can easily derive  that
contribution to the decoherence factor coming from low energy modes excited by 
the quench is given by 
\begin{equation}
\hat D(t) \approx \exp\left(-N\Upsilon(t)/  \tau_Q^{r\nu/(1+z\nu)}\right).
\label{general}
\end{equation}
Above $\tau_Q^{r\nu/(1+z\nu)}$ 
captures the universality-class dependence of decoherence caused by 
non-equilibrium crossing of the quantum critical point, while
$\Upsilon(t)$ stands for the non-universal (system-dependent) response to driving.  
Moreover, $r$ is the environment  dimensionality and $N$ is the 
environment size (number of spins/particles/etc.).
Note also that the contribution of the  momentum modes that were not excited by the 
quench adds another factor besides $\hat D$ to the full decoherence factor $D$
in complete  analogy to what we have seen for the Ising environment in Eqs. (\ref{L_approx}) and (\ref{L_approx1}).
We expect that this ``adiabatic'' factor is well approximated by instantaneous 
ground state fidelity as was shown to be the case in the Ising model. 

\noindent Finally, we mention that 
 (\ref{general}) reveals also that 
non-equilibrium decoherence is expected to be exponential in the number of topological 
defects created by the quench:
$$
N/\hat\xi^r\sim N/\tau_Q^{r\nu/(1+z\nu)}.
$$
Note that $\hat\xi$ 
provides a typical distance between topological defects created by the quench.

\section{Conclusions}
We have demonstrated rich manifestations of
 decoherence induced by a dynamical  phase transition in the environment.
In particular, we have shown that near the critical point decoherence is 
dramatically enhanced not just by the increased susceptibility, but
also by the excitations that depend on the past history of the system.
Our results provide a strong indication of a 
universality-class dependent behavior that bears remarkable similarity 
to the dynamics of defect formation in non-equilibrium phase transitions
\cite{1kzm, Znature}. 

This work should stimulate future  theoretical  studies.
In particular, it would be interesting to study  both universal and system-dependent contributions 
to decoherence dynamics in other out-of-equilibrium environments,  especially 
those described by a different universality class than the Ising model. 
It would be also interesting to find out if a similar
description can be worked out for systems whose wave function is not given by the 
product over momentum modes. Generalization of our results to finite temperatures
can  provide yet another  extension of our work.

We expect that our findings will 
be experimentally accessible in future cold atom/ion quantum simulators 
of condensed matter systems \cite{LewensteinReviewMisc}. Moreover, 
it should be also possible to extend them to optomechanical systems where a cold atom cloud
undergoing a QPT (the environment) is coupled to a
mirror (a quantum system) \cite{Santos2010}. Finally, generalizing the idea from 
\cite{Goldstein2010}, the non-equilibrium
enhancement of the decoherence rate of the  qubit  might find its applications
in magnetometry.  

\section{Acknowledgements}

This work is supported by U.S. Department of Energy through the LANL/LDRD Program. 
We acknowledge stimulating discussions with Rishi Sharma, and thank Jacek Dziarmaga
for his comments on the  manuscript.

\appendix

\renewcommand{\thesection}{}
\section{Simplification of the exact solution}
\renewcommand{\theequation}{A\arabic{equation}}
Exact solutions (\ref{exact}) and (\ref{exact1}) for the Bogolubov modes
can be simplified  for (i) modes evolving adiabatically; 
(ii) small momentum modes excited by crossing the critical point
at $g_c=1$; and (iii) for large momentum modes excited due to 
non-adiabatic crossing of the critical point at $g_c=-1$. 
The resulting expressions can be used to simplify decoherence factor in 
momentum space: $F_k$. Their knowledge allows for accurate determination of the
decoherence factor $D$ away from the critical points. 
We will briefly outline below how  compact expressions 
for $F_k$ can be derived.

{\bf Adiabatic evolution of modes:} 
Modes evolving adiabatically are described by well-known ground state solutions 
taken at {\it instantaneous} value of the magnetic field $g(t)$ \cite{sachdev,Jacek2005}.
This implies that 
$$
v_k^\pm=\cos(\theta^\pm/2), \ \ u_k^{\pm}=\sin(\theta^\pm/2),
$$
where $\theta^\pm\in[0,\pi]$, 
$\cos(\theta^\pm) = \epsilon_\pm/\sqrt{1+\epsilon_\pm^2}$,
and $\epsilon_\pm = (-g(t)\mp\delta+\cos k)/\sin k$. 
The decoherence factor in  momentum 
space takes the compact form
\begin{equation}
F_k=\cos^2\left(\frac{\theta^++\theta^-}{2}\right),
\label{swinka}
\end{equation}
which can be simplified by the Taylor expansion
$$
F_k=1-\frac{\delta^2\sin^2k}{(1-2g\cos k+g^2)^2} + {\cal O}(\delta^4).
$$

{\bf Small momentum expansion for the modes excited after crossing the critical 
point at $g_c=1$:}
We assume here that $g(t)<1$ and consider only $k\ll\pi/4$ relevant for slow
transitions ($\tau_Q\gg1$).
Our expansion of (\ref{exact}) and (\ref{exact1}) is based on large $|iz_\pm|$ expansion of the 
Weber functions. 
We define $z_\pm=\alpha_\pm\exp(i\pi/4)$, where
$$
\alpha_\pm(t) = 2\sqrt{\tau_Q}(-g(t) \mp\delta+\cos k).
$$
To expand Weber functions we
assume below that $\alpha_\pm\gg1$, which implies that 
$$
\tau_Q\gg\frac{1}{4(1-g(t)-\delta)^2}
$$
for small $k$ modes. This condition can be equivalently 
rewritten to 
$$
1-g(t)\gg \delta, {\cal O}(1)/\sqrt{\tau_Q}.
$$
To proceed, we need the following identities \cite{Whittaker}.
For $|{\rm arg}(s)|<3\pi/4$ 
$$
{\cal D}_m(s)= e^{-s^2/4}s^m \left[1+{\cal O}(s^{-2})\right],
$$
while for $\pi/4<{\rm arg}(s)<5\pi/4$ 
$$
{\cal D}_m(s)= e^{im\pi}{\cal D}_m(-s)+\frac{\sqrt{2\pi}}{\Gamma(-m)}e^{i(m+1)\pi/2}
{\cal D}_{-m-1}(-is).
$$
Keeping only the leading 
terms in the expansion of the Weber functions in (\ref{exact}) and (\ref{exact1}), we find that 
\begin{eqnarray}
v_k^\pm(t) &\approx& \sqrt{\frac{\pi\tau_Q'}{2}}
\frac{
e^{-\pi\tau_Q'/8}e^{-i[\alpha_\pm^2(t)+\tau_Q'\ln\alpha_\pm(t)]/4}
}{\Gamma(1-i\tau_Q'/4)}, 
\label{uv_simplified}
\end{eqnarray}
\begin{eqnarray}
u_k^\pm(t) &\approx& e^{-\pi\tau_Q'/4}
e^{i\pi/4}e^{i[\alpha_\pm^2(t)+\tau_Q'\ln\alpha_\pm(t)]/4}.
\label{uv_simplified1}
\end{eqnarray}
Having these expressions we get
\begin{eqnarray}
v_k^{+*}(t)v_k^{-}(t) =  \left[1-\exp\left(-\pi\tau_Q'/2\right)\right]\exp\left(-i\chi(t)\right),
\label{uv_prod}
\end{eqnarray}
\begin{eqnarray}
u_k^{+*}(t)u_k^{-}(t) = \exp\left(-\pi\tau_Q'/2\right)\exp\left(i\chi(t)\right),
\label{uv_prod1}
\end{eqnarray}
where 
\begin{equation}
\chi(t) = [\alpha^2_-(t)-\alpha^2_+(t)]/4 + \tau_Q'\ln[\alpha_-(t)/\alpha_+(t)]/4.
\label{chi}
\end{equation}
\noindent Next we additionally simplify these results by considering 
$k\ll\pi/4$ and $\delta\ll1$ limits. 
These assumptions result in $\tau_Q'\approx4k^2\tau_Q$ and 
$\chi(t)\approx 4\delta\tau_Q[1-g(t)]$.
Substituting $g(t)=1-t/\tau_Q$  into (\ref{chi}) one gets 
$\chi(t)\approx4t\delta$.
After some
algebra these expressions 
allow for writing the decoherence factor in  momentum 
space  as 
\begin{equation}
F_k(t) \approx 1-4\sin^2\left(4t\delta\right)
\left(e^{-2\pi\tau_Qk^2}-e^{-4\pi\tau_Qk^2}\right).
\label{fk_quench}
\end{equation}

{\bf Large momentum expansion for the modes excited after crossing the critical 
point at $g_c=-1$:} We assume here that $g(t)<-1$.
The calculations are similar as above with the only difference that 
slow driving through the critical point at $g_c=-1$
excites momentum modes near $k=\pi$ rather then those close to $k=0$.
The small $k$ modes, excited during crossing the critical point at $g_c=1$,
are not affected by driving through the critical point at $g_c=-1$:
(\ref{uv_simplified}) - (\ref{fk_quench}) hold  for $g(t)<-1$.

\noindent 
The solutions (\ref{uv_simplified}) and (\ref{uv_simplified1}) are valid again, but the expansion now requires 
$$
\tau_Q\gg\frac{1}{4(-1-g(t)-\delta)^2},
$$
because we need to substitute $\cos k\approx-1$ into $\alpha_\pm$.
Equations (\ref{uv_prod}) and (\ref{uv_prod1}) are the same, but the phase $\chi$ 
and modified quench time scale $\tau_Q'$
have different expansions for $\delta\ll1$ and $\pi-k\ll\pi/4$:
$\chi(t)\approx-4\tau_Q\delta(1+g)$ and $\tau_Q'\approx4(k-\pi)^2\tau_Q$.

\noindent Wrapping it up we obtain
$$
F_k(t)\approx1-4
\left(e^{-2\pi\tau_Q(k-\pi)^2}-e^{-4\pi\tau_Q(k-\pi)^2}\right)
\sin^2\eta(t),
$$
where $\eta(t)=4\tau_Q\delta(1+g(t))$.

\end{document}